\documentclass[aps,prl,twocolumn,floats,amsmath,amssymb,showpacs,showkeys,preprintnumbers,floatfix]%
{revtex4-1}

\usepackage{graphicx}
\usepackage{dcolumn}
\usepackage{color}

\begin{document}

\title{Large deviations and rain showers}

\author{ Michael Wilkinson}
\affiliation{Department of Mathematics and Statistics,
The Open University, Walton Hall, Milton Keynes, MK7 6AA, England,}

\begin{abstract}

Rainfall from ice-free cumulus clouds requires collisions 
of large numbers of microscopic droplets to create every raindrop. 
The onset of rain showers can be surprisingly rapid, much faster than the mean time
required for a single collision. Large-deviation theory is used to explain
this observation.

\end{abstract}

%\pacs{92.60.Nv}{Cloud physics and chemistry}
%\pacs{92.60.hk}{Convection, turbulence, and diffusion}
%\pacs{92.60.Mt}{Particles and aerosols}

\maketitle

The dynamics of the onset of rainfall from ice-free (\lq warm') cumulus clouds
is poorly understood \cite{Mas71,Pru+97,Sha03}. 
A rain drop grows by collisions of microscopic water droplets.
A large number of microscopic droplets must combine to make one
rain drop: the volume increase is a factor of approximately 
one million. The collision rates in the early stages of the growth process
are low (typically of order one collision per hour). Given the large 
number of collisions 
which must occur, it is very hard to understand the observation that 
rain showers can be initiated 
over relatively short periods, of perhaps twenty minutes. 

One possible resolution is a consequence of the large number 
of microscopic droplets which must combine to make a raindrop.
This implies that only very few drops are required to undergo explosive 
growth, and perhaps there are sufficient rare combinations of rapid multiple
collisions to explain rainfall: this point has previously been emphasised 
by Kostinski and Shaw \cite{Kos+05}. A quantitative approach is required
to show whether the large number of collisions required for runaway 
growth can occur with a sufficiently high probability. Because this problem involves 
the analysis of rare events, methods based upon large deviation 
theory \cite{Fre+84,Tou09} are used in this Letter to investigate the hypothesis 
that rare combinations of rapid collisions trigger showers. 
It is shown that a rain shower can develop
over a timescale which is a small fraction of the mean timescale for one collision.

First, consider some observations and estimates 
\cite{Mas71,Pru+97,Sha03} which illustrate the difficulties in 
making a quantitative description of rainfall. 
A convecting cumulus cloud which could produce showers may have 
droplets of mean radius radius is $a_0=10\,\mu {\rm m}$, which result from 
condensation onto aerosol nuclei. Raindrops have a much larger
size, typically $1\,{\rm mm}$. The volume of a droplet which becomes 
a raindrop therefore increases by a very large factor, denoted by ${\cal N}$, which 
is typically ${\cal N}\approx 10^6$.
The number density of microscopic droplets is typically of order $N_0=2.5\times 10^8\,{\rm m}^{-3}$,
which gives a liquid water content, expressed as a volume fraction, $\Phi_{\rm l}\approx 
4\pi N_0a_0^3/3\approx 10^{-6}$.
The cloud depth may be $h=2\times 10^3\, {\rm m}$ and the typical vertical 
velocity of air inside the cloud has magnitude
$U\approx 2\,{\rm m}\,{\rm s}^{-1}$, so that the turnover time for convection is
approximately $\tau_h=10^3\,{\rm s}$. Rainfall from this type of cloud can 
develop over a timescale of approximately $20\,{\rm min}
\approx 10^3\,{\rm s}$. 

Droplets which undergo a geometrical collision (the impact parameter is less than the sum
of the radii) might not coalesce, because the streamlines of small droplets curve around larger
ones. In fact, if the Navier-Stokes equations were a complete description, droplets would never
collide, because there would always be a lubricating film of air between them.
The coalescence efficiencies $\varepsilon$ of small droplets are somewhat uncertain, but it
is widely accepted that they are low for typical cloud droplets \cite{Mas71,Pru+97}.
If the larger droplet has radius below $20\,\mu{\rm m}$, it is believed that $\varepsilon \le 0.1$,
and that for radius $10\,\mu {\rm m}$, $\varepsilon \le 0.03$ \cite{Pru+97}. For droplets of size
$a=50\,\mu {\rm m}$ colliding with droplets of size $a=10\,\mu {\rm m}$, however,
the efficiencies are expected to be close to unity \cite{Pru+97,Mas71}.

Collisions between droplets settling at a different rate yield a very small
collision rate. The Stokes law for the drag on a sphere at low Reynolds number
indicates that the gravitational settling rate is
\begin{equation}
\label{eq: 1}
v=\tau_{\rm p}g=\kappa a^2
\ ,\ \ \
\kappa=\frac{2}{9}\frac{\rho_{\rm l}}{\rho_{\rm g}}\frac{g}{\nu}
\end{equation}
where $\tau_{\rm p}$ is the response time characterising the Stokes
drag on a droplet, $\rho_{\rm l}$ is the density of liquid water, 
and $\rho_{\rm g}$ and $\nu$ are, respectively, the density
and kinematic viscosity of air.
Inserting values for air and water at $5^\circ {\rm C}$ gives
$\kappa\approx 1.4\times 10^8\,{\rm m}^{-1}{\rm s}^{-1}$, 
so that when $a_0=10\,\mu{\rm m}$ the terminal velocity is
$v\approx 1.4\times 10^{-2}\,{\rm ms}^{-1}$ and the response time is $1.4\times 10^{-3}\,{\rm s}$.
The collision rate of a drop of radius $a_1$ with a gas
of droplets of radius $a_0$ is
\begin{equation}
\label{eq: 2}
R_1=\pi \varepsilon N_0 (a_0+a_1)^2 \kappa (a_1^2-a_0^2)
\end{equation}
where $\varepsilon$ is the collision efficiency.
Setting $a_1-a_0=2.5\,\mu{\rm m}$ and $\varepsilon=0.03$ in 
addition to the parameters defined above
gives $R_1\approx  2\times 10^{-5}{\rm s}^{-1}$. The rate of coalescence
of typical sized water droplets due to collisions is therefore very small.

Cumulus clouds are turbulent because of convective instability.
Saffman and Turner \cite{Saf+56} investigated the role of turbulence in facilitating collisions
between water droplets. In the case of very small droplets, the collision rate due
to turbulence is a consequence of shearing motion. The shear rate of small-scale
motions in turbulence is the inverse of the Kolmogorov timescale, 
$\tau_{\rm K}=\sqrt{\nu/\epsilon}$, where $\epsilon $ is the rate of dissipation
per unit mass. According to the Saffman-Turner model, shear induces a collision speed 
of order $a_0/\tau_{\rm K}$. They argue that the corresponding collision rate is
\begin{equation}
\label{eq: 3}
R_{\rm turb}=\sqrt{\frac{8\pi}{15}}\frac{N_0\varepsilon (2a)^3}{\tau_{\rm K}}
\ .
\end{equation}
For the parameters of the cloud model, the rate of dissipation 
is $\epsilon\sim U^2/\tau_h\approx 2\times 10^{-3}\,{\rm m}^2{\rm s}^{-3}$, giving 
$\tau_{\rm K}\approx 70\,{\rm ms}$, which gives $R_{\rm turb}\approx 10^{-6}\,{\rm s}^{-1}$,
which is negligible. The effects of turbulence are dramatically increased
when the effects of droplet inertia are significant: this was noticed 
in numerical experiments by Sundaram and Collins \cite{Sun+97},
who ascribed the effect to a clustering effect termed \lq preferential 
concentration' \cite{Max87}. More recent work has proposed an alternative 
mechanism, which has been termed the \lq sling effect' \cite{Fal+02}, and which has
been explained in terms of the existence of caustics in the velocity 
field of the droplets \cite{Wil+06}. Inertial effects are measured
by the Stokes number, ${\rm St}\equiv\tau_{\rm p}/\tau_{\rm K}$. Recent 
numerical studies \cite{Vos+14} (see also \cite{Gra+13}) show that the collision rate
is greatly enhanced by effects due to caustics for ${\rm St}>0.3$, equation  
(\ref{eq: 3}) is a good estimate when ${\rm St}\ll 1$. 
While it is in principle possible for turbulence to be responsible for an enhanced 
collision rate of water droplets due to inertial effects, the parameters of the cloud 
model discussed above yield 
${\rm St}\approx 2\times 10^{-2}$, where there is no significant enhancement.
While there is a consensus that turbulence is important for the formation of rain showers 
\cite{Bod+10}, turbulent enhancement of collision rates does not appear to be sufficient. 

Now consider how to model the onset of a shower.
It has already been remarked that showers
occur on a timescale which may be smaller than the typical 
timescale for one collision. It is, therefore, reasonable
to assume that the runaway droplets are falling through a background of 
droplets which have not yet coalesced. As a runaway droplet falls it
collides with a large number ${\cal N}$ of small droplets of size $a_0$. 
The time between successive collisions may be assumed to be independent Poisson 
processes. If the time between the collision with index $n$ and the 
previous collision is $t_n$, the time for a droplet to experience 
runaway growth is 
\begin{equation}
\label{eq: 4}
T=\sum_{i=1}^{\cal N} t_n
\end{equation}
where the $t_n$ are independent random variables with a Poisson 
distribution
\begin{equation}
\label{eq: 5}
P_n(t_n)=R_n\exp(-R_n t_n)
\ .
\end{equation}
The problem is to determine the statistics of $T$ in the limit as ${\cal N}\to \infty$. 
The rates for successive collisions increase as the size of the falling drop grows.
Because all of the collision rates $R_n$ scale in the same way as a function of the 
droplet size $a_0$ and the number density $N_0$,  write
\begin{equation}
\label{eq: 6}
R_n=R_1 f(n)
\ .
\end{equation}
Here $R_1$ depends upon the properties of the cloud but the function $f(n)$ 
is the same for all clouds.
In order to identify the form of $f(n)$, consider the rate of collision 
of a large droplet resulting from $n$ previous collisions with a gas
of small droplets of radius $a_0$. The radius of the large droplet 
is $a_n=n^{1/3}a_0$. When $n$ is large it may assumed that the 
collision efficiency is $\varepsilon \approx 1$ and $a_n\gg a_0$
so that $R_n\sim \pi N_0 \kappa a_n^4\propto n^{4/3}$, 
which suggests setting $f(n)=n^{4/3}$. However during 
the early stages of droplet growth, the collision efficiency for the 
first few collisions is small, but increases rapidly with $n$.
In what follows $f(n)$ is assumed to be a power-law
\begin{equation}
\label{eq: 7}
R_n=R_1 n^\gamma
\ .
\end{equation}
If the collision efficiency of droplets were unity, it would be 
appropriate to set $\gamma=4/3$. 
Because the collision efficiency 
of droplets at the crucial initial stage of their growth is small, the collision
rate increases more rapidly as the size of the falling droplet increases.
When the droplets are between $10\,\mu{\rm m}$ and $50\,\mu{\rm m}$ it is 
reasonable to model the product of the collision rate and the collision efficiency as
being proportional to $a^6$, that is to $n^2$, where $n$ is the number of collisions \cite{Kos+05}. 
In other cases, such as solid precipitation (snow), other values of $\gamma$ may be appropriate.   
In the following $\gamma $ is left as an adjustable parameter, but special consideration 
is given to $\gamma =2$, because it gives a good approximation to terrestrial rainfall,
and to $\gamma=4/3$, because this may be a good approximation for atmospheres 
on other planets where the collision efficiency might not limit the rate of coalescence.

It is necessary to determine  the probability density for the time $T$ being a 
very small fraction of its mean value, $\langle T\rangle$. Inspired by large deviation theory
\cite{Fre+84,Tou09}, the probability density of $T$ is written in an exponential form:
\begin{equation}
\label{eq: 8}
P(T)=\frac{1}{\langle T\rangle}\exp[-J(\tau)]
\ ,\ \ \ \ 
\tau=\frac{T}{\langle T\rangle}
\ .
\end{equation}
When $f(n)$ is given by (\ref{eq: 7}), the mean time for explosive growth
converges as ${\cal N}\to \infty$ when $\gamma>1$:
\begin{equation}
\label{eq: 9}
\lim_{{\cal N}\to \infty}\langle T\rangle=
\lim_{{\cal N}\to \infty}\frac{1}{R_1}\sum_{n=1}^{\cal N} \frac{1}{f(n)}=
\frac{1}{R_1}\zeta(\gamma)
\ .
\end{equation}
where $\zeta$ is the Riemann zeta function.
The function $J(\tau)$ in (\ref{eq: 8}) is often termed the entropy 
in texts on large deviation theory. It will be necessary to determine the entropy 
function $J(\tau)$ from the rate function $f(n)$. 

After a drop has grown to a size where it is much larger than the typical droplets,
and where the collision efficiency is approximately unity, it falls rapidly and collects other droplets
in its path. Consider a drop of size $a_1$ falling through a \lq gas' of much smaller droplets,
with liquid volume fraction $\Phi_{\rm l}$. The larger drop falls with velocity $v=\kappa a_1^2$ and 
grows in volume at a rate $\pi \varepsilon a_1^2 \Phi_{\rm l}v=4\pi a_1^2\dot a_1$, where $\dot a_1$
is the rate of increase of the drop radius. The rate of increase 
of the radius of the \lq collector' drop as a function of the distance $x$ through which it has fallen is
\begin{equation}
\label{eq: 10}
\frac{{\rm d}a}{{\rm d}x}=\frac{\varepsilon \Phi_{\rm l}}{4}
\ .
\end{equation}
Note that this expression is valid whether or not the terminal velocity 
is given by the small Reynolds number approximation, (\ref{eq: 1}).
In the case of droplets which reach a radius of approximately $1\,{\rm mm}$, 
the collision efficiency $\varepsilon$ is close to unity throughout most of the fall. 
The droplet radius after falling through a cloud of depth $h$ is
therefore $a(h)\sim \Phi_{\rm l}h/4$.
It will be assumed that the most relevant  collector drops are those that
started at the top of the cloud, so that the volumetric growth factor is 
\begin{equation}
\label{eq: 11}
{\cal N}=\left(\frac{h}{4a_0}\right)^3\Phi_{\rm l}^3
\ .
\end{equation}
Using the representative values given above gives
${\cal N}\approx 10^5$. 

The rate of change of the liquid water content of a cloud due to the runaway 
growth of droplets is
\begin{equation}
\label{eq: 12}
\frac{{\rm d}\Phi_{\rm l}}{{\rm d}t}=-\Phi_{\rm l} {\cal N} P(t)
\ .
\end{equation}
Note that the growth factor ${\cal N}$ and the probability density 
for runaway growth after time $t$ are both functions of $\Phi_{\rm l}$, but if 
the objective is to understand the onset of a rain shower it suffices to evaluate 
these quantities with the initial value $\Phi_{\rm l}(0)$. Using (\ref{eq: 8}) 
for $P(t)$, the condition for the timescale $t^\ast$ where there is a significant reduction 
in $\Phi_{\rm l}(t)$ is ${\cal N} \exp[-J(t^\ast/\langle T\rangle)]=1$, 
that is
\begin{equation}
\label{eq: 13}
t^\ast=\tau^\ast \langle T\rangle
\ ,\ \ \ \ 
J(\tau^\ast)={\rm ln}\, {\cal N}
\ .
\end{equation}
The droplet volume growth factor 
$ {\cal N}$ was estimated in equation (\ref{eq: 11}). 
To determine  determine the solution of (\ref{eq: 13}) for $t^\ast$, 
it is necessary to determine the entropy function $J(\tau)$ for the random sum defined by
(\ref{eq: 4}) and (\ref{eq: 5}).

\begin{figure}
\centering
\includegraphics[width=0.45\textwidth]{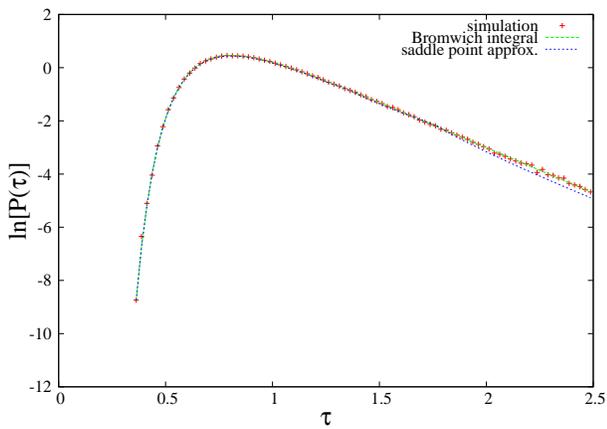}
\caption{Plot of ${\rm ln}[P(\tau)]$, for ${\cal N}=4000$ and 
$\gamma=4/3$. The results of simulation, evaluation of the Bromwich integral, and the saddle point approximation are in excellent agreement.}
\label{fig: 1}
\end{figure}

\begin{figure}
\centering
\includegraphics[width=0.45\textwidth]{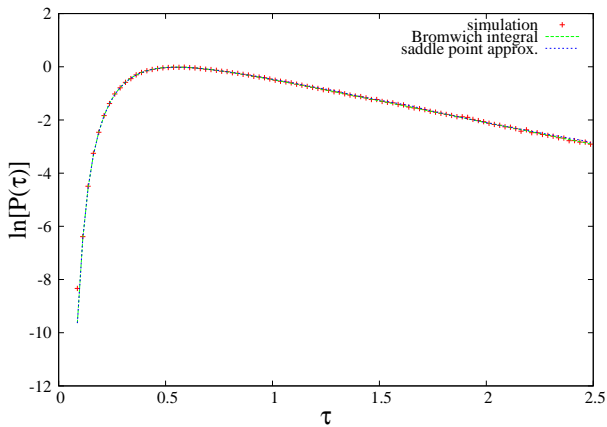}
\caption{Plot of ${\rm ln}[P(\tau)]$, for ${\cal N}=4000$ and 
$\gamma=2$.}
\label{fig: 2}
\end{figure}

To compute $J(T)$ consider a cumulent generating function
$\lambda(k)$, defined by writing
\begin{equation}
\label{eq: 14}
\exp[-\lambda (k)]=\langle \exp(-kT)\rangle=\int_0^\infty {\rm d}T\ P(T)\,\exp(-kT)
\ .
\end{equation}
Because the $t_n$ are independent, with a distribution given by (\ref{eq: 5}):
\begin{equation}
\label{eq: 15}
\lambda(k)=\sum_{n=1}^{\cal N} {\rm ln}\langle \exp(-kt_n)\rangle
=\sum_{n=1}^{\cal N} {\rm ln}\left(1+\frac{k}{R_n}\right)
\ .
\end{equation}
Now consider how to obtain $P(T)$ from $\lambda(k)$. Noting that $\exp[-\lambda(k)]$ 
is the Laplace transform of $P(T)$, application of the Bromwich integral formula 
for inversion gives    
\begin{equation}
\label{eq: 16}
P(T)=\frac{1}{2\pi {\rm i}}\int_{C -{\rm i}\infty}^{C+{\rm i}\infty}
{\rm d}z\ \exp[zT-\lambda(z)]
\end{equation}
where $C >-R_1$.
The integral is dominated by contributions from the neighbourhood of a saddle 
at $z=k^\ast$, where 
\begin{equation}
\label{eq: 17}
T=\sum_{n=1}^{\cal N} \frac{1}{R_n+k^\ast}
\end{equation}
which is to be solved for $k^\ast$ given a value of $T$. The probability density
$P(T)$ is then approximated by
\begin{equation}
\label{eq: 18}
P(T)=\frac{1}{R_1}\frac{1}{\sqrt{2\pi {\cal S}(k^\ast)}}\exp[-J(\tau)]
\end{equation}
where $\tau=T/\langle T\rangle$ and ${\cal S}(k)$ is the magnitude of the second derivative 
of the exponent in (\ref{eq: 16}). 
Equations (\ref{eq: 17}) and (\ref{eq: 18}) cannot be solved exactly and 
explicitly for $J(\tau)$. Consider how
to write down a parametric representation of $J(\tau)$ using a scaled variable, $\kappa$, 
defined by $\kappa=k^\ast/R_1$. The dimensionless 
time for raindrop formation is
\begin{equation}
\label{eq: 19}
\tau(\kappa)=\sum_{n=1}^{\cal N} \frac{1}{\kappa +n^\gamma}
\left[{\sum_{n=1}^{\cal N} n^{-\gamma}}\right]^{-1}
\end{equation}
and the entropy function is
\begin{equation}
\label{eq: 20}
J(\kappa)=\sum_{n=1}^{\cal N} {\rm ln}\left(1+\kappa n^{-\gamma}\right)
-\sum_{n=1}^{\cal N} \frac{\kappa n^{-\gamma}}{1+\kappa n^{-\gamma}}
\ .
\end{equation}
Figure \ref{fig: 1} shows the distribution of $\tau=T/\langle T\rangle $ for the case $\gamma =4/3$, 
with ${\cal N}=4000$ and $R_1=1$, comparing the results of simulation of (\ref{eq: 4}), 
the Bromwich integral (\ref{eq: 16}),
the saddle-point approximation, equations (\ref{eq: 18}), 
(\ref{eq: 19}), (\ref{eq: 20}), which are all in excellent agreement.
Figure \ref{fig: 2} makes a similar comparison for the case $\gamma=2$,
which is most relevant to rain showers. 
In both cases the entropy function increases very rapidly as $\tau\to 0$, indicating 
that the value of $\tau^\ast=t^\ast/\langle T\rangle$ is quite 
insensitive to the value of ${\rm ln}\,{\cal N}$. 
It is clear from figure \ref{fig: 2} that
the solution of equation (\ref{eq: 13}) gives small values of $\tau^{\ast}$
when ${\cal N}$ is large. Numerical evaluation of (\ref{eq: 16})
with $\gamma=2$ gives $\tau^{\ast}\approx 0.077$ when ${\cal N}=10^5$ 
and $\tau^{\ast}\approx 0.068$ when ${\cal N}=10^6$. Alternatively, 
in terms of $\langle t_1\rangle=\langle T\rangle/\zeta(\gamma)$, when $\gamma=2$, the 
predicted time for onset of a shower is a small fraction of the mean time for the 
first collision: 
$t^{\ast}\approx 0.128\langle t_1\rangle$ when ${\cal N}=10^5$ 
and $t^{\ast}\approx 0.112\langle t_1\rangle$ when ${\cal N}=10^6$.

It is clear from figures \ref{fig: 1} and \ref{fig: 2} that the entropy increases 
very rapidly as $\tau\to 0$, and it is desirable to find asymptotic behaviour of $J(\tau)$ 
when $\tau \ll 1$. This limit corresponds to $\kappa\gg 1$. 
In the limit where ${\cal N}\to \infty$ and $\kappa\gg 1$ in equations 
(\ref{eq: 19}) and (\ref{eq: 20}), the sum 
in the numerator of (\ref{eq: 19}) is approximated 
by an integral:
\begin{equation}
\label{eq: 21}
\tau(\kappa)\sim \frac{1}{\zeta(\gamma)}\int_0^\infty {\rm d}n\ \frac{1}{\kappa+n^\gamma}
=\kappa^{\frac{1}{\gamma}-1}F(\gamma)
\end{equation}
with
\begin{equation}
\label{eq: 22}
F(\gamma)=\frac{1}{\gamma \zeta(\gamma)}\int_0^\infty {\rm d}x\ \frac{x^{\frac{1}{\gamma}-1}}{1+x}
\ .
\end{equation}
A similar approach applied to (\ref{eq: 20}) gives
\begin{equation}
\label{eq: 23}
J(\kappa)=\kappa^{\frac{1}{\gamma}}G(\gamma)
\ ,\ \ \ 
G(\gamma)=\frac{1}{\gamma}\int_0^\infty {\rm d}x\ \frac{{\rm ln}(1+x)}{x^{\frac{1}{\gamma}+1}}
-\zeta(\gamma)F(\gamma)
\ .
\end{equation}
Eliminating $\kappa$ from (\ref{eq: 22}) an (\ref{eq: 23}) shows that 
in the limit as $\tau\to 0$ the entropy function has a power-law divergence: 
to leading order
\begin{equation}
\label{eq: 24}
J(\tau)\sim G(\gamma)[F(\gamma)]^{\frac{1}{\gamma-1}} \tau^{-\frac{1}{\gamma-1}}
\equiv K(\gamma)\tau^{-\frac{1}{\gamma-1}}
\ .
\end{equation}
This indicates that the probability density has a non-algebraic singularity as 
$\tau\to 0$: $P(\tau)\sim \exp(-K/\tau)$ when $\gamma=2$, or $P(\tau)\sim \exp(-K/\tau^3)$
when $\gamma=\tfrac{4}{3}$.  Numerical integration gives $K(\tfrac{4}{3})\approx 0.8799$, 
whereas $K(2)=\pi^2/4\zeta(2)=\tfrac{3}{2}$  exactly.

The conclusion is that rain showers can commence in a timescale
which is short compared to the mean time for the first collision between 
droplets, with the timescale for onset being approximately one eighth of the mean time 
for first collision in the case of the more realistic model ($\gamma=2$). 
Thus large deviation theory has resolved an apparent 
paradox of meteorology, that rain showers can start very quickly, on timescale
which are short compared to typical mean collision times.
 
This calculation does not resolve all of the uncertainties about initiation 
of rain showers. Clouds can exist 
for a long period without producing a rain shower, before depositing 
a large fraction of their water content over a short time. Shower activity 
is associated with convective motion in clouds, and it has been suggested 
that turbulence facilitates collisions. For typical levels of turbulence, 
however, turbulent enhancement of collisions does not appears to be 
sufficient. It seems as if non-collisional mechanisms involving 
convection must may a role in initiating the cascade \cite{Wil14}.

This work was initiated during a visit to the Kavli Institute for Theoretical 
Physics, Santa Barbara, where this research was supported in part by the 
National Science Foundation under Grant No. NSF PHY11-25915.

\end{document}